\documentclass{epl}%
\usepackage{amsfonts}
\usepackage{amsmath}
\usepackage{amssymb}
\usepackage{graphicx}
\usepackage{caption2}%
\providecommand{\U}[1]{\protect \rule{.1in}{.1in}}
\renewcommand \caption{}

\institute{
\inst{1} Department of Microelectronics and Information Technology, The Royal
Institute of Technology, KTH, Electrum 229, SE-164 40 Kista, Sweden\\
\inst{2} Department of Physics, Tsinghua University, Beijing 100084, China
} 
\begin{document}

\title{Improved practical decoy state method in quantum key distribution with
parametric down conversion source}
\author{Qin Wang\inst{1}
\and Xiang-Bin Wang\inst{2}
\and Gunnar Bj\"{o}rk\inst{1}
\and Anders Karlsson\inst{1}}
\maketitle

\begin{abstract}
In this paper, a new decoy-state scheme for quantum key distribution with
parametric down-conversion source is proposed. We use both three-intensity
decoy states and their triggered and nontriggered components to estimate the
fraction of single-photon counts and quantum bit-error rate of single-photon,
and then deduce\ a more accurate value of key generation rate. The final key
rate over transmission distance is simulated, which shows that we can obtain a
higher key rate than that of the existing methods, including our own earlier work.

\end{abstract}

\section{Introduction}

Quantum key distribution has attracted extensive attentions for its
unconditional security compared with conversional cryptography
\cite{benn,maye,shor,eker,deut,deut2}. However, there still exist several
technical limitations in practice, such as imperfect single-photon sources,
large loss channels and inefficient detectors, which will impair the security.
Fortunately, many methods have been devised to deal with these imperfect
conditions \cite{scran,kko,gott,hwan,wang,lo}, among which, decoy-state method
is thought to be a very useful candidate for substantially improving the
performance of QKD.

Decoy-state method was firstly proposed by Hwang \cite{hwan}, and advanced by
Wang and Lo \textit{et al. }\cite{wang,lo,wang2,ma1,harr}\textit{ }assuming a
weak coherent source (WCS). Subsequently, it was extended to parametric down-
conversion sources (PDCS) \cite{hori,maue,qin1}. The main idea of decoy-state
method is to randomly change the intensity of each pulse among different
values, which allows one to estimate the behavior of vacuum, single-photon and
multi-photon states individually. As a result, Eve's presence will be
detected. Recently, more and more interesting ideas have been put forward to
improve the performance of QKD \cite{adac,qi,maue}, such as the one by Adachi
\textit{et al. }\cite{adac}.\textit{ }In their proposal, both triggered and
nontriggered components of PDCS are used to do some estimations for final
secure key, and it needs only one intensity to transmit. However, because the
intensity cannot be changed during the whole experiment, and dark counts
cannot be measured directly, then the worst case of their contribution must be
considered, which will inevitably limit final key rate and transmission distance.

In this paper, we propose a new practical decoy-state scheme with PDCS, in
which not only three decoy states with different intensities ($0,\mu
,\mu^{\prime}$), but also all their triggered and nontriggered components are
used to estimate the lower bound of fraction of single photon counts ($Y_{1}$)
and upper bound of quantum bit-error rate (QBER) of single-photon ($e_{1}$).
As a result, a more accurate value of key generation rate, compared with
existing methods, can be obtained.

\section{Improved decoy state method}

In our new scheme, we can essentially use almost the same experimental setup
as that in our previous proposal \cite{qin1}, except that Bob's detector need
to work no matter what Alice's detector is triggered or not.

As is well known, the state of two-mode field from PDCS\bigskip \ is
\cite{yurk,lutk}:%
\begin{align*}
\left \vert \Psi \right \rangle _{TS}  &  =\overset{\infty}{\underset{n=0}%
{\sum \sqrt{P_{n}}}}\left \vert n\right \rangle _{T}\left \vert n\right \rangle
_{S},\\
P_{n}  &  =\frac{x^{n}}{\left(  1+x\right)  ^{n+1}},
\end{align*}
where $\left \vert n\right \rangle $ represents an $n$-photon state, and $x$ is
the intensity (average photon number) of one mode. Mode T (trigger) is
detected by Alice, and mode S (signal) is sent out to Bob. We request Alice to
randomly change the intensity of her pump light among three values, so that
the intensity of one mode is randomly changed among $0,\mu,\mu^{\prime}$ (and
$\mu<\mu^{\prime}$).

We denote $q_{n}$ as the probability of triggering at Alice's detector when an
$n$-photon state is emitted,%

\begin{equation}
q_{n}=1-\left(  1-\eta_{A}\right)  ^{n},\text{ \  \ }n=1,2,3...
\end{equation}
where $\eta_{A}$ is the detecting efficiency at Alice's side, then the
nontriggering probability is $\left(  1-q_{n}\right)  .$ We define $Y_{n}$ to
be the yield of an $n$-photon state, i.e., the probability that Bob's detector
clicks whenever Alice sends out state $|n\rangle$; we also define $Q_{n}$ be
the gain of a $n$-photon state, i.e., the rate of events when Alice emits
$n$-photon state and Bob detects the signal, which can be divided into two
groups, triggered by Alice $Q_{n}^{(t)}$, and the rest $Q_{n}^{(ut)}$; and
$Q_{x}$ be the overall rate according to intensity $x$, ($x$ can be $0,\mu
,\mu^{\prime}$), it can also be divided into two groups, triggered by Alice
$Q_{x}^{(t)}$, and the rest $Q_{x}^{(ut)}$, which can be expressed as:%

\begin{align}
Q_{x}^{(t)}  &  =Y_{0}\frac{d_{A}}{1+x}+%
{\displaystyle \sum_{i=1}^{\infty}}
Y_{n}\left[  1-\left(  1-\eta_{A}\right)  ^{n}\right]  \frac{x^{n}}{\left(
1+x\right)  ^{n+1}},\\
Q_{x}^{(ut)}  &  =Y_{0}\frac{1-d_{A}}{1+x}+%
{\displaystyle \sum_{i=1}^{\infty}}
Y_{n}\left(  1-\eta_{A}\right)  ^{n}\frac{x^{n}}{\left(  1+x\right)  ^{n+1}},
\end{align}
where $d_{A}$ is the dark count rate of Alice's detector.

In the next step, we will use the triggered events of $\mu$ ($Q_{\mu}^{(t)}$)
and the nontriggered events of $\mu^{\prime}$ ($Q_{\mu^{\prime}}^{(ut)}$) to
deduce a tight bound of the fraction of single-photon counts ($Y_{1}$).%
\begin{align}
Q_{\mu}^{(t)}  &  =Y_{0}\frac{d_{A}}{1+\mu}+%
{\displaystyle \sum_{i=1}^{\infty}}
Y_{n}\left[  1-\left(  1-\eta_{A}\right)  ^{n}\right]  \frac{\mu^{n}}{\left(
1+\mu \right)  ^{n+1}},\\
Q_{\mu^{\prime}}^{(ut)}  &  =Y_{0}\frac{1-d_{A}}{1+\mu^{\prime}}+%
{\displaystyle \sum_{i=1}^{\infty}}
Y_{n}\left(  1-\eta_{A}\right)  ^{n}\frac{{\mu^{\prime}}^{n}}{\left(
1+\mu^{\prime}\right)  ^{n+1}}.
\end{align}
The two equations lead to:%
\begin{align}
&  \frac{(1+\mu)\left(  \frac{\mu^{\prime}}{1+\mu^{\prime}}\right)  ^{2}%
Q_{\mu}^{(t)}}{1-\left(  1-\eta_{A}\right)  ^{2}}-\frac{(1+\mu^{\prime
})\left(  \frac{\mu}{1+\mu}\right)  ^{2}Q_{\mu^{\prime}}^{(ut)}}{\left(
1-\eta_{A}\right)  ^{2}}\nonumber \\
&  =Y_{0}\left[  \frac{\left(  \frac{\mu^{\prime}}{1+\mu^{\prime}}\right)
^{2}d_{A}}{1-(1-\eta_{A})^{2}}-\frac{\left(  \frac{\mu}{1+\mu}\right)
^{2}(1-d_{A})}{(1-\eta_{A})^{2}}\right] \nonumber \\
&  +Y_{1}\left[  \frac{\eta_{A}}{1-(1-\eta_{A})^{2}}\frac{\mu}{1+\mu}\left(
\frac{\mu^{\prime}}{1+\mu^{\prime}}\right)  ^{2}-\frac{1}{1-\eta_{A}}\frac
{\mu^{\prime}}{1+\mu^{\prime}}\left(  \frac{\mu}{1+\mu}\right)  ^{2}\right]
+\nonumber \\
&  +\sum_{n=3}^{\infty}Y_{n}\left[  \frac{1-\left(  1-\eta_{A}\right)  ^{n}%
}{1-(1-\eta_{A})^{2}}\frac{\mu^{n}{\mu^{\prime}}^{2}}{(1+\mu)^{n}%
(1+\mu^{\prime})^{2}}-\left(  1-\eta_{A}\right)  ^{n-2}\frac{{\mu^{\prime}%
}^{n}{\mu}^{2}}{\left(  1+\mu^{\prime}\right)  ^{n}(1+\mu)^{2}}\right]  .
\label{complex1}%
\end{align}
Assuming the condition%
\[
\frac{1-\left(  1-\eta_{A}\right)  ^{n}}{1-(1-\eta_{A})^{2}}\frac{\mu^{n}%
{\mu^{\prime}}^{2}}{(1+\mu)^{n}(1+\mu^{\prime})^{2}}-\left(  1-\eta
_{A}\right)  ^{n-2}\frac{{\mu^{\prime}}^{n}{\mu}^{2}}{\left(  1+\mu^{\prime
}\right)  ^{n}(1+\mu)^{2}}\leq0
\]
can be satisfied, i.e.
\begin{equation}
\mu \leq \frac{a\mu^{\prime}}{1+\mu^{\prime}-a\mu^{\prime}},
\end{equation}
where $a=\left(  \frac{\left(  1-\eta_{A}\right)  ^{n-2}-\left(  1-\eta
_{A}\right)  ^{n}}{1-\left(  1-\eta_{A}\right)  ^{n}}\right)  ^{\frac{1}{n-2}%
},$ (because the values of $\mu$ and $\mu^{\prime \text{ }}$can be chosen
independently, the assumption above can be easily satisfied in experiment,)
then Eq. (7) leads to the following inequality:
\begin{equation}
Y_{1}\geq Y_{1}^{L}=\frac{\frac{(1+\mu)\left(  \frac{\mu^{\prime}}%
{1+\mu^{\prime}}\right)  ^{2}Q_{\mu}^{(t)}}{1-\left(  1-\eta_{A}\right)  ^{2}%
}-\frac{(1+\mu^{\prime})\left(  \frac{\mu}{1+\mu}\right)  ^{2}Q_{\mu^{\prime}%
}^{(ut)}}{\left(  1-\eta_{A}\right)  ^{2}}-Y_{0}\left[  \frac{\left(
\frac{\mu^{\prime}}{1+\mu^{\prime}}\right)  ^{2}d_{A}}{1-(1-\eta_{A})^{2}%
}-\frac{\left(  \frac{\mu}{1+\mu}\right)  ^{2}(1-d_{A})}{(1-\eta_{A})^{2}%
}\right]  }{\frac{\eta_{A}}{1-(1-\eta_{A})^{2}}\frac{\mu}{1+\mu}\left(
\frac{\mu^{\prime}}{1+\mu^{\prime}}\right)  ^{2}-\frac{1}{1-\eta_{A}}\frac
{\mu^{\prime}}{1+\mu^{\prime}}\left(  \frac{\mu}{1+\mu}\right)  ^{2}}.
\end{equation}
This gives rise to the gain of single-photon pulse for triggered and
nontriggered components as:%
\begin{align}
Q_{1}^{(t)}\left(  x\right)   &  =Y_{1}\eta_{A}\frac{x}{(1+x)^{2}},\\
Q_{1}^{(ut)}\left(  x\right)   &  =Y_{1}\left(  1-\eta_{A}\right)  \frac
{x}{(1+x)^{2}},
\end{align}
and $x$ may be $\mu$ or $\mu^{\prime}$ here. Also, if we have observed the
quantum bit-error rate (QBER) for triggered and nontriggered pulses of
intensity $x$, $E_{x}^{(t)}$, $E_{x}^{(ut)}$, we can upper bound the QBER
value of single-photon pulse as:%
\begin{align}
e_{1}  &  \leq \frac{(1+x)^{2}E_{x}^{(t)}Q_{x}^{(t)}-(1+x)Y_{0}d_{A}/2}%
{Y_{1}\eta_{A}x}=e_{a},\\
e_{1}  &  \leq \frac{(1+x)^{2}E_{x}^{(ut)}Q_{x}^{(ut)}-(1+x)Y_{0}(1-d_{A}%
)/2}{Y_{1}(1-\eta_{A})x}=e_{b},
\end{align}
Combing the two bounds, we have:%
\begin{equation}
e_{1}^{U}=\min \left \{  e_{a},e_{b}\right \}  .
\end{equation}
Normally, we use the value from $x=\mu$ for a tight estimation of $e_{1}$.
Given all these, we can use the following formula to calculate the final
key-rate of triggered signal pulses \cite{gott}:
\begin{equation}
R^{(t)}\geq \frac{1}{2}\left \{  -Q_{\mu^{\prime}}^{(t)}f\left(  E_{\mu^{\prime
}}^{(t)}\right)  H_{2}\left(  E_{\mu^{\prime}}^{(t)}\right)  +Q_{0}%
^{(t)}+Q_{1}^{(t)}\left[  1-H_{2}\left(  e_{1}\right)  \right]  \right \}  ,
\end{equation}
where the factor of $\frac{1}{2}$ comes from the cost of basis mismatch in
Bennett-Brassard 1984 (BB84) protocol; $f(E_{\mu^{\prime}})$ is a factor for
the cost of error correction given existing error correction systems in
practice. We assume $f=1.22$ here \cite{bras}. $H_{2}\left(  x\right)  $ is
the binary Shannon information function, given by
\[
H_{2}\left(  x\right)  =-x\log_{2}(x)-(1-x)\log_{2}(1-x).
\]

Furthermore, if the transmission distance is not so large, the nontriggered
component can also be used to generate secret key just as in Adachi \textit{et
al}'s proposal \cite{adac}:%
\begin{align}
R^{(both)}  &  \geq \frac{1}{2}\{-Q_{\mu^{\prime}}^{(t)}f\left(  E_{\mu
^{\prime}}^{(t)}\right)  H_{2}\left(  E_{\mu^{\prime}}^{(t)}\right)
-Q_{\mu^{\prime}}^{(ut)}f\left(  E_{\mu^{\prime}}^{(ut)}\right)  H_{2}\left(
E_{\mu^{\prime}}^{(ut)}\right) \nonumber \\
&  +Q_{0}^{(t)}+Q_{0}^{(ut)}+\left(  Q_{1}^{(t)}+Q_{1}^{(t)}\right)  \left[
1-H_{2}\left(  e_{1}\right)  \right]  \}.
\end{align}
In this case the final key rate is given by: $R=\max \left \{  R^{(t)}%
,R^{(both)}\right \}  $.

\section{Numerical simulation}

In an experiment, we need to observe the values of $Q_{0}^{(t)}$, $Q_{\mu
}^{(t)}$, $Q_{\mu^{\prime}}^{(t)}$, $Q_{0}^{(ut)}$, $Q_{\mu}^{(ut)}$,
$Q_{\mu^{\prime}}^{(ut)}$ and $E_{\mu}^{(t)}$, $E_{\mu^{\prime}}^{(t)}$,
$E_{\mu}^{(ut)}$, $E_{\mu^{\prime}}^{(ut)}$, and then deduce the lower bound
of fraction of single-photon counts ($Y_{1}$) and upper bound QBER of
single-photon pulses ($e_{1}$) by theoretical results, and then one can
distill the secure final key. In order to make a faithful estimation, we need
a channel model to forecast what values for $Q_{0}^{(t)}$, $Q_{\mu}^{(t)}$,
$Q_{\mu^{\prime}}^{(t)}$, $Q_{0}^{(ut)}$, $Q_{\mu}^{(ut)}$, $Q_{\mu^{\prime}%
}^{(ut)}$ and $E_{\mu}^{(t)}$, $E_{\mu^{\prime}}^{(t)}$, $E_{\mu}^{(ut)}$,
$E_{\mu^{\prime}}^{(ut)}$ \emph{would} be, if we \emph{did} the experiment
without Eve in principle.

Suppose $\eta$ is the combined overall transmittance and detection efficiency
between Alice and Bob; $t_{AB}$ is the transmittance between Alice and Bob,
$t_{AB}=10^{-\alpha L/10}$; $\eta_{B}$ is the transmittance in Bob's side,
$\eta=t_{AB}.\eta_{B}$. Following these assumptions, $Y_{n}=d_{B}%
+1-(1-\eta)^{n}$, which approximates $1-(1-\eta)^{n}$ when $n\geqslant1$, and
the observed value for $Q_{x}^{(t)}$ and $Q_{x}^{(ut)}$ should be:%
\begin{align}
Q_{x}^{(t)}  &  =\frac{d_{A}d_{B}}{1+x}+%
{\displaystyle \sum_{i=1}^{\infty}}
\left[  d_{B}+1-\left(  1-\eta \right)  ^{n}\right]  \left[  1-\left(
1-\eta_{A}\right)  ^{n}\right]  \frac{x^{n}}{\left(  1+x\right)  ^{n+1}},\\
Q_{x}^{(ut)}  &  =\frac{\left(  1-d_{A}\right)  d_{B}}{1+x}+%
{\displaystyle \sum_{i=1}^{\infty}}
\left[  d_{B}+\left(  1-\eta \right)  ^{n}\right]  \left(  1-\eta_{A}\right)
^{n}\frac{x^{n}}{\left(  1+x\right)  ^{n+1}},
\end{align}
where $x$ can be $\mu$ or $\mu^{\prime}$, and $d_{B}$ is the dark count rate
of Bob's detectors.

We use the following for the error rate of an \emph{n-photon} state
\cite{lo}:
\begin{equation}
e_{n}=\frac{e_{0}d_{B}+e_{d}[1-(1-\eta)^{n}]}{d_{B}+1-(1-\eta)^{n}},
\end{equation}
where $e_{0}=1/2$, $e_{d}$ is the probability that the survived photon hits a
wrong detector, which is independent of the transmission distance. Below we
shall assume $e_{d}$ to be a constant. Therefore, the observed $E_{x}%
^{(t)},E_{x}^{(ut)}$ values should be:
\begin{align}
E_{x}^{(t)}  &  =\frac{%
{\displaystyle \sum_{n=0}^{\infty}}
e_{n}Q_{n}^{(t)}\left(  x\right)  }{%
{\displaystyle \sum_{n=0}^{\infty}}
Q_{n}^{(t)}\left(  x\right)  },\\
E_{x}^{(ut)}  &  =\frac{%
{\displaystyle \sum_{n=0}^{\infty}}
e_{n}Q_{n}^{(ut)}\left(  x\right)  }{%
{\displaystyle \sum_{n=0}^{\infty}}
Q_{n}^{(ut)}\left(  x\right)  }.
\end{align}

\begin{figure}[ptb]
\begin{center}
\includegraphics[scale=0.6]{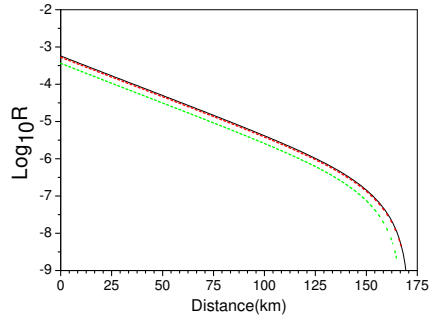}
\end{center}
\caption{Fig. 1 (Color online) Final key rates vs transmission distance for
decoy-state method. The solid line is the ideal result where the fraction of
single-photon counts and QBER of single-photon pulses are known exactly; the
dotted lines and dashed lines are the simulation results with finite
decoy-state method, among which, the upper line is our new result using only
triggered events with $\mu=\frac{a\mu^{\prime}}{1+\mu^{\prime}-a\mu^{\prime}}%
$; the lower line is the result of our previous proposal with $\mu=0.1$,
($\mu^{\prime}$ has the optimal value at each point in each line.)}%
\label{Fig1}%
\end{figure}

\begin{figure}[ptb]
\begin{center}
\includegraphics[scale=0.6]{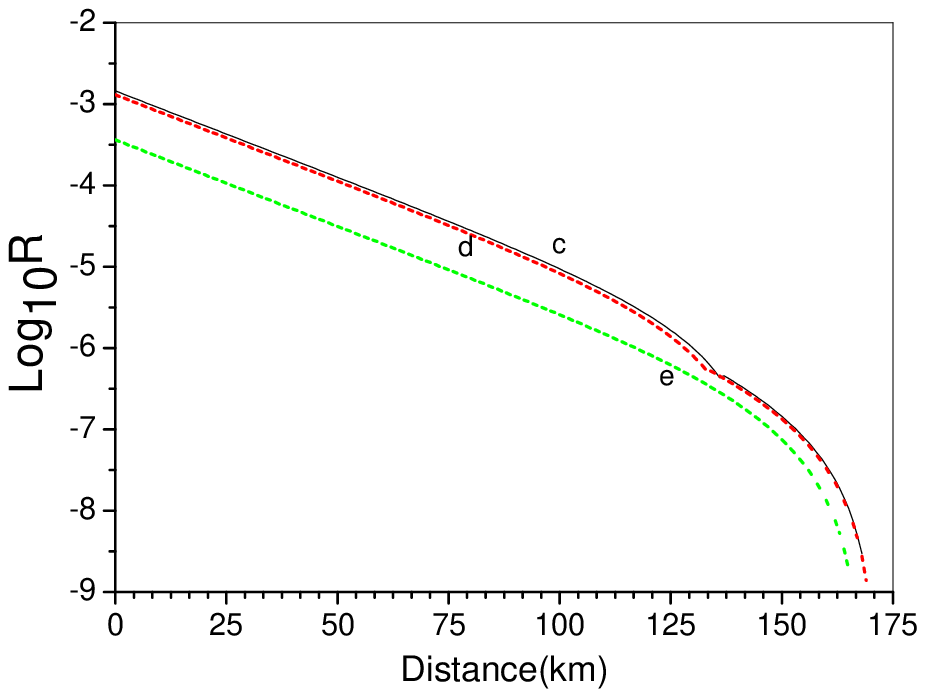}
\end{center}
\caption{Fig. 2 (Color online) Final key rates vs transmission distance for
decoy-state method. The solid line is the ideal result where the fraction of
single-photon counts and QBER of single-photon pulses are known exactly; the
dotted lines and dashed lines are the simulation results with finite
decoy-state method, among which, the upper line is our new result using both
triggered and nontriggered events with $\mu=\frac{a\mu^{\prime}}{1+\mu
^{\prime}-a\mu^{\prime}}$; the lower line is the result of our previous
proposal with $\mu=0.1$, ($\mu^{\prime}$ has the optimal value at each point
in each line.)}%
\label{Fig2}%
\end{figure}

\begin{figure}[ptb]
\begin{center}
\includegraphics[scale=0.6]{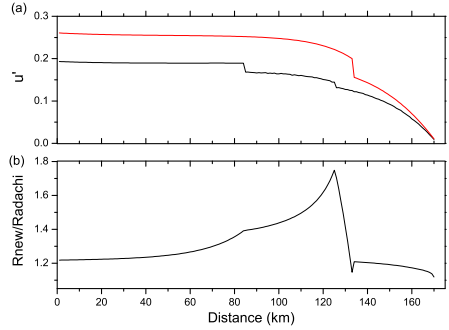}
\end{center}
\caption{Fig. 3 (Color online) (a) The optimal value of $\mu^{\prime}$ vs
transmission distance. The upper line is the result of our new proposal (
$\mu=\frac{a\mu^{\prime}}{1+\mu^{\prime}-a\mu^{\prime}}$), and the lower line
is the result of Adachi \textit{et al}. (b) The ratio of key rates between our
new proposal and Adachi \textit{et al}'s vs transmission distance.}%
\label{Fig3}%
\end{figure}

In practical implementation of QKD, we often use non-degenerated
down-conversion to produce photon pairs \cite{dani,thew,mori}, with one photon
at the wavelength convenient for detection acting as heralding signal, and the
other at the telecommunication windows for optimal propagation along the fiber
or in open air acting as heralded signal. We can now calculate the final key
rate with the assumed values above. For convenience of comparing with the
results of Adachi \textit{et al}. \cite{adac}, we use the same parameters as
used in their paper which mainly come from Gobby, Yuan and Shields (GYS)
experiment \cite{gobb}. At Alice's side, $d_{A}\bigskip=10^{-6},\eta_{A}=0.5$;
at Bob's side, $d_{B}\bigskip=1.7\times10^{-6},\eta_{B}=0.045,e_{d}=0.033;$
and the channel loss is $\alpha=0.21(dB/km).$ At each distance we choose the
optimal value for $\mu^{\prime}$, so that we can have the highest key rate,
and the final results are shown in Fig. 1, 2, 3 (according to Eq. (7),
$\mu=\frac{a\mu^{\prime}}{1+\mu^{\prime}-a\mu^{\prime}}$ is chosen in our new proposal).

Fig. 1 shows the key generation rate against transmission distance compared
with our previous results \cite{qin1}, (only triggered events are used.) It
shows that our new scheme can generate a higher key rate than the old one even
using only triggered signal.

Fig. 2 shows the key generation rate against transmission distance compared
with our previous results \cite{qin1}, (both triggered and nontriggered events
are used.) From it we can see that our new results can approach the ideal
values very closely. Moreover, there is no need to use a quite weak decoy
state or nontriggered signal. For example, at the distance of 50 km, setting
$\mu_{opt}^{\prime}=0.255,\mu=0.113,\eta_{A}=0.5$ in our new scheme, and
$\mu_{opt}^{\prime}=0.143,\mu=0.113,\eta_{A}=0.5$ in the old one, we can get a
ratio of key rate between the two scheme as 3.8. The reason that we can get a
more accurate estimation of key rate ($R$) in the new proposal is as follows:
we don't omit those high order items (in formula (6)) in the deduction of
$Y_{1}$, but use them to deduce a relationship between the intensity of decoy
state ($\mu$) and signal state ($\mu^{\prime}$), which inevitably results in a
more bound estimation of $Y_{1}.$(In addition, there is an inflexion in curve
c (d) at the distance about 134 km, because the nontriggered events cease to
contribute to the key rate.)

Fig. 3 (a) shows the optimal values of $\mu^{\prime}$ in our new proposal
(setting $\mu=\frac{a\mu^{\prime}}{1+\mu^{\prime}-a\mu^{\prime}}$) and those
in Adachi \textit{et al}'s; Fig. 3 (b) shows the ratio of the key generation
rate between our new scheme and Adachi \textit{et al}'s against transmission
distance, (both triggered events and nontriggered events are used.) It shows
that our result is always larger than theirs when using almost the same level
of data size \cite{explain}.

From the figures above, we can see that, our new results are better than those
of both our previous proposal and Adachi \textit{et al}'s. As is known
\cite{ma1}, to give a more accurate estimation of the key rate, the value of
$\mu$ should be chosen to be the smaller the better. In our previous proposal,
the key rate could also be very close to the ideal value given a very weak
decoy state $\mu$. However, in a practical experiment, considering statistical
errors, $\mu$ cannot be too weak. So in our new scheme, we deduce a relation
between $\mu$ and $\mu^{\prime}$, and at each point, both $\mu$ and
$\mu^{\prime}$ can be chosen with optimal values, which results in a more
accurate estimation. Comparing with Adachi \textit{et al}'s proposal, the
advantages of our proposal are as follows: Firstly, dark counts can be
measured directly; secondly, a weaker decoy state $\mu$ is used to get a more
accurate estimation of $Y_{1\text{ }}$and $e_{1}$, and a stronger signal of
$\mu^{\prime}$ is used to get a higher secure key rate.

\section{Conclusion}

In summary, we have proposed a new decoy-state scheme in QKD with PDCS, in
which we use both three-intensity decoy-states and their triggered and
nontriggered components to get a tight bound of the fraction of single-photon
counts and single-photon QBER, This allows us to accurately deduce the value
of key generation rate. Finally, the key generation rate vs transmission
distance is numerically simulated. The simulations show that our new results
are better than those of the existing proposal. Furthermore, our proposal only
assumes existing experimental technology, which makes the scheme a practical
candidate in the implementation of QKD. \acknowledgments

Q. W thanks Yoritoshi Adachi (Osaka University) and S$\acute{e}$bastien Sauge
for useful discussions, and Gao-Xin Qiu for help in numerical simulation. This
work was funded by the European Community through the QAP (Qubit
Applications-015848) project, and also supported by Chinese National
Fundamental Research Program and Tsinghua Bairen Program.

\end{document}